\documentclass[11pt, final]{article}
\usepackage{etex}
\usepackage{a4wide}
\usepackage{enumitem}
\usepackage{amssymb}
\usepackage{algorithmic}
\usepackage{algorithm}
\usepackage{amsmath,euscript}
\usepackage{array}
\usepackage[all,cmtip]{xy}
\usepackage{graphicx}
\usepackage{float}
\usepackage{mathtools}
\usepackage{subcaption}
\usepackage{todonotes}
\usepackage{booktabs}
\usepackage{cite}
\usepackage{PaperDef}
\newcommand{\proof}{\par\noindent{\bf Proof}. \ignorespaces}

\newcommand {\rank}       {\mathop{\rm rank}\nolimits}

\makeatletter
\def\revddots{\mathinner{\m
kern1mu\raise\p@
    \vbox{\kern7\p@\hbox{.}}\mkern2mu
    \raise4\p@\hbox{.}\mkern2mu\raise7\p@\hbox{.}\mkern1mu}}
\makeatother
\mathcode`@="8000 
{\catcode`\@=\active\gdef@{\mkern1mu}}
\numberwithin{equation}{section}
\numberwithin{table}{section}
\numberwithin{figure}{section}
\newtheorem{theorem}{Theorem}[section]

\begin{document}
\topmargin = -5ex;

\thispagestyle{empty}
\begin{center}

\renewcommand{\thefootnote}{\fnsymbol{footnote}}

\begin{minipage}{0.29\textwidth}
	\includegraphics[width=1in]{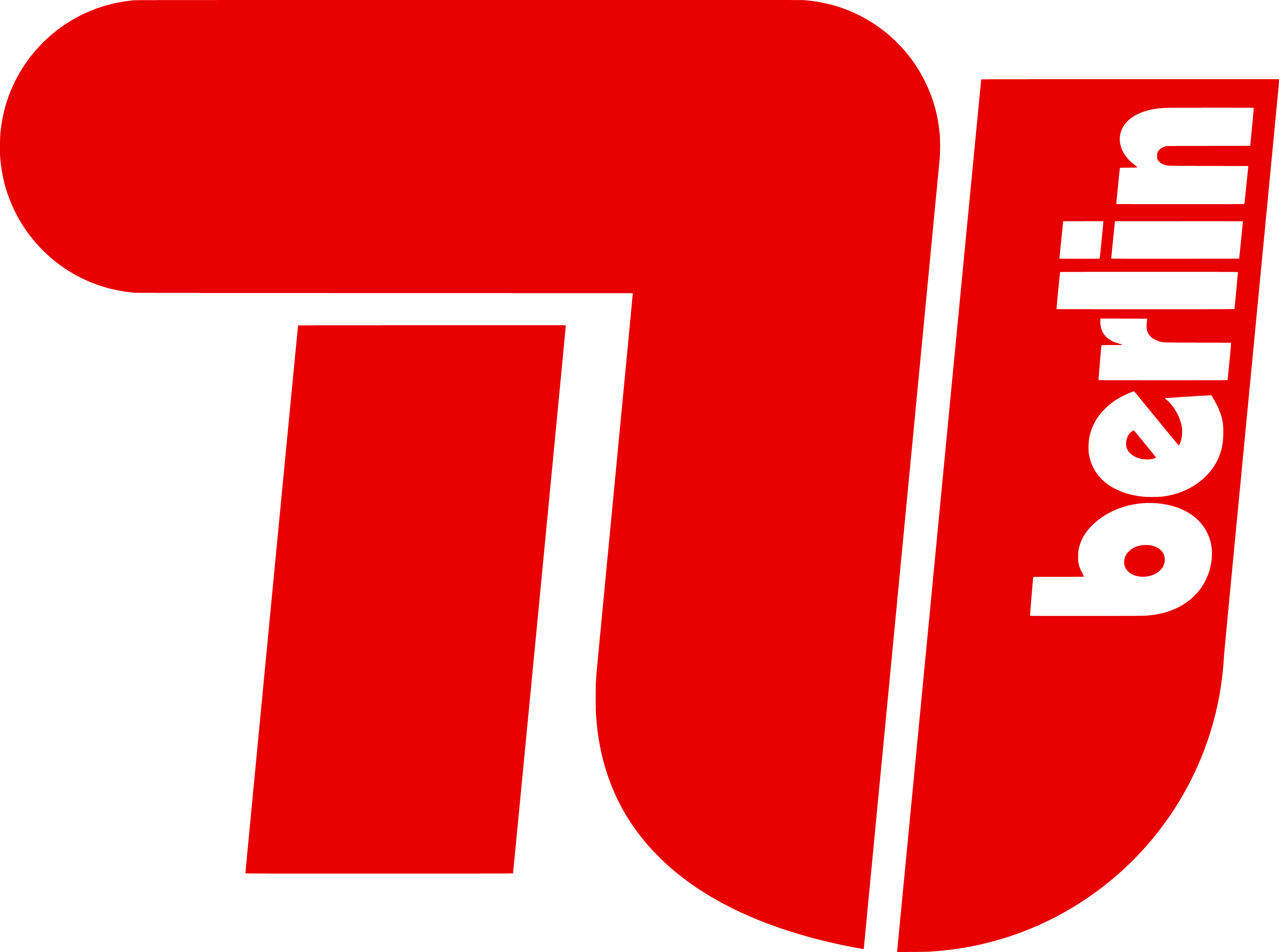}
\end{minipage}
\begin{minipage}{0.69\textwidth}
	\begin{flushright}
		{\Huge Technische Universit\"at Berlin}
		\Large Institut f\"ur Mathematik
	\end{flushright}
\end{minipage}
\\ [50mm]
\scalebox{0.95}{\bf\LARGE Model Reduction for  Large-scale Dynamical Systems} \\{\bf  \LARGE with Inhomogeneous Initial Conditions}\\[13mm]
{\large\bf \begin{tabular}{ccccc}Christopher A. Beattie & \qquad & Serkan Gugercin& \qquad &Volker Mehrmann \end{tabular}} \\ [18mm]
{\bf Preprint 15-2016} \\
\vfill {\bf Preprint-Reihe des Instituts f\"ur Mathematik} \\[1mm]
{\bf Technische Universit\"at Berlin} \\[1mm]
{\bf \tt http://www.math.tu-berlin.de/preprints} \\[8mm]
{\bf Preprint 15-2016 \hfill August 2016}
\end{center}
\newpage
~
\newpage
\setcounter{page}{1}

\title{Model Reduction for  Systems with \\ Inhomogeneous Initial Conditions}
\author{Christopher Beattie \footnotemark[1]\ \footnotemark[2]
\and Serkan Gugercin \footnotemark[1]\ \footnotemark[6]
\and Volker Mehrmann \footnotemark[2] \footnotemark[3]
}
\maketitle

\begin{abstract}
We consider the model reduction problem for linear time-invariant dynamical systems having nonzero (but otherwise indeterminate) initial conditions. Building upon the observation that the full system response is decomposable as a superposition of the response map for an unforced system having nontrivial initial conditions and the response map for a forced system having null initial conditions,  we develop a new approach that involves reducing these component responses independently and then combining the reduced responses into an aggregate reduced system response.
This approach allows greater flexibility and offers better approximation properties than other comparable methods.

\end{abstract}

\vskip .3truecm
\noindent
{\bf AMS(MOS) subject classification:} 34H05, 65L70, 65L05, 37-04

\maketitle
\renewcommand{\thefootnote}{\fnsymbol{footnote}}

\footnotetext[1]{
	Department of Mathematics, Virginia Tech, Blacksburg, VA,  \texttt{$\{$beattie,gugercin$\}$@vt.edu}.
}
\footnotetext[2]{
	Institut f\"ur Mathematik, MA 4-5 TU Berlin, Str. d. 17. Juni 136, D-10623 Berlin, Germany,  \texttt{mehrmann@math.tu-berlin.de}.
}
\footnotetext[3]{The authors gratefully acknowledge the support by the Deutsche
Forschungsgemeinschaft (DFG) as part of the collaborative research center SFB 1029 Substantial efficiency increase in gas turbines through direct use of coupled unsteady combustion
and flow dynamics, project A02 {\it Development of a reduced order model of pulsed detonation
combustor}  }
\footnotetext[5]{
	The work of this author was supported in part by the Einstein Foundation Berlin. 	
}
\footnotetext[6]{
	The work of this author was supported in part by the Alexander von Humboldt Foundation.
}
\renewcommand{\thefootnote}{\arabic{footnote}}

\section{Introduction}
We consider model reduction for linear time-invariant dynamical systems having \emph{nonzero} initial conditions and a
state-space realization given as
\begin{eqnarray} \nonumber
\dot{\bx}(t) &=& \bA \bx(t)  + \bB \bu(t) ,\quad \bx(0) = \bx_0 \\
 {\by}(t)  &=& \bC \bx(t), \label{eq:FOM}
\end{eqnarray}
where $\bA \in \IR^{n\times n}$, $\bB \in \IR^{n\times m}$, and $\bC \in \IR^{p\times n}$.  For each $t\geq0$,
$\bx(t) \in \IR^n$, $\bu(t) \in \IR^m$, and $\by(t) \in \IR^p$ are, respectively, the \emph{states}, \emph{inputs}, and \emph{outputs} of \eqref{eq:FOM}.   $\bx_0 \in \IR^n$ is the (generally) nonzero initial condition, prescribed at $t_0=0$.  The choice of initial time $t_0=0$ is arbitrary and a treatment of general initial times follows immediately by translation. We assume throughout that \eqref{eq:FOM} is \emph{asymptotically stable}, so that all eigenvalues of $\bA$ have (strictly) negative real parts.

Our goal is to construct a reduced order model having the same form, denoted here as
\begin{eqnarray} \nonumber
\dot{\tbx}(t) &=& \tbA \tbx(t)  + \tbB \bu(t) ,\quad \tbx(0) = \tbx_{0} \\
 \tby(t)  &=& \tbC \tbx(t),\label{eq:ROM}
\end{eqnarray}
where $\tbA \in \IR^{r\times r}$, $\tbB \in \IR^{r\times m}$, and $\tbC \in \IR^{p\times r}$ have been chosen
for some $r \ll n$,
together with a reduced initial condition, $\tbx_{0} \in \IR^r$, such that the reduced output
$\tby(t)$ approximates $\by(t)$ well over a wide range of inputs, $\bu(t)$, and initial conditions, $\bx_0$.

For null initial conditions ($\bx_0 = \mathbf{0}$), there already exists a wide range of available methods for constructing high-fidelity reduced models.  Such approaches include Lyapunov-based methods such as Balanced Truncation (\BT) \cite{MulR76,Moo81} and optimal Hankel Norm Approximation \cite{Glo84}, interpolatory methods such as the Iterative Rational Krylov Algorithm (\IRKA) \cite{GugAB08} and spectral zero interpolation \cite{Ant05a,Sor05}, and data-driven methods such as the Loewner framework \cite{MayA07} and Vector Fitting \cite{GusS99,DrmGB14}.
Detailed discussion of many of these techniques with comparative advantages and disadvantages may be found for example, in  \cite{Ant05,AntBG10,BauBF14,BeaG15}.

Circumstances change when the initial condition is \emph{nonzero}, and one natural approach (see e.g., \cite{BauBF14}) considers a translated state vector, $\widehat{\bx}(t) = \bx(t)- \bx_0$.   One may rewrite (\ref{eq:FOM}) in terms of  $\widehat{\bx}(t)$, which now has a null initial condition, $\widehat{\bx}(0)=0$, and this allows the use of reduction methods designed for null initial conditions.   The reduced state trajectory is translated back to recover an approximation of the original state trajectory resulting from the nontrivial initial condition (see for example, \cite{CarFCA13}).   This can be an effective strategy if there is a single initial condition that is known {\it a priori}.   Observe that the transient (though possibly significant) effect of a nontrivial initial condition has been transformed into an asymptotic state bias that is likely to persist in the final reduced model even after the final reverse translation.   Thus, this approach may overemphasize the effect of the initial condition and so, the associated reduced models may produce poor response approximations.  Indeed, reduced models produced in this way should not be expected to yield good approximations if different initial conditions are used from that used in the original reduction.

To improve this situation, a new approach was suggested in  \cite{HeiRA11} that creates a reduced model providing a good approximation to the true output, $\by(t)$, not only for a variety of input functions, $\bu(t)$, but also for a variety of initial conditions
$\bx_0$, making the model reduction process largely independent of the specification of particular initial conditions.
For this approach, the initial conditions are not presumed to be known \emph{a priori}, but are assumed to lie in a known $n_0$-dimensional subspace, $\mathcal{X}_0$, spanned by the columns of a matrix $\bX_0 \in \IR^{n \times n_0}$.
 The method of \cite{HeiRA11} proceeds by appending the basis $\bX_0$ to the input-to-state matrix,  $\bB$ and then performs balanced truncation using the augmented input-to-state matrix, $[\bB~~\bX_0]$.  It is shown in \cite{HeiRA11} that this approach can significantly outperform regular balanced truncation, which would not have a mechanism to take $\bx_0$ into account.

We follow a  somewhat different strategy here in incorporating initial condition information into the model reduction process.
Our approach is based on the simple observation that the output $\by(t)$ of the linear dynamical system \eqref{eq:FOM} is a superposition of two outputs, one that corresponds to the response of  \eqref{eq:FOM} to $\bu(t)$ with $\bx(0) = 0$
and a second one associated with $\bu(t) = 0$ and $\bx(0) = \bx_0$.
This leads us to two independent model reduction processes, one on the input-to-output map ($\Suy$) and one on the initial condition-to-output map ($\Sxy$).

There are several  advantages that accrue with this approach. First of all, it allows complete flexibility on the accuracy and order of the reduced models corresponding to each of $\Suy$ and  $\Sxy$.  This flexibility can be significant since one of the two maps, $\Suy$ or $\Sxy$, might be significantly harder to reduce than the other.   Indeed, considering independent model reduction processes on each of $\Suy$ and  $\Sxy$ allows one to consider different model reduction techniques for each of the two maps (this will be demonstrated in Section~\ref{sec:main}) and leads us also to aggregate error bounds that clearly reveal the contributions resulting from the $\Suy$ approximation and the $\Sxy$ approximation.  This lends additional flexibility in controlling and balancing errors corresponding to the two components. Note that working instead with the augmented input-to-state matrix $[\bB~~\bX_0]$ as in \cite{HeiRA11}, can mask discrepancies in the difficulty of reducing $\Suy$ and  $\Sxy$ (examples will be provided in Section~\ref{sec:results}) and forces one to adopt a fixed method and single reduction order for the combined mapping.
Considering the reduction of $\Suy$ and $\Sxy$ independently also permits greater flexibility in retaining significant underlying system structure, such as port-Hamiltonian structure \cite{BeaG11,GugPBS12,PolS10}.
 Finally, in the context of model reduction for descriptor systems (differential-algebraic systems), where initial conditions must be chosen consistently, this requirement persists also in the reduced model, and so makes the independent treatment of initial conditions essential.   Themes related to descriptor/differential-algebraic systems will not be pursued here but will be discussed in a separate paper.

The  paper is organized as follows. In Section~\ref{sec:prep} we briefly review different model reduction techniques such as Balanced Truncation, the Iteratively Rational Krylov Algorithm,  as well as the augmented  method of \cite{HeiRA11}.  In Section~\ref{sec:main} we introduce the new methodology together with error bounds. We present various numerical examples in Section~\ref{sec:results} followed by conclusions in Section~\ref{sec:conc}.

\section{Background on Selected Methods for Model Reduction} \label{sec:prep}
\subsection{Methods for homogeneous initial conditions}
We briefly recall the classical methods of Balanced Truncation (\BT) and the Iterative Rational Krylov Algorithm (\IRKA), both of which will play a crucial role in the proposed model reduction framework presented in Section~\ref{sec:main}.
\subsubsection{Balanced truncation (\BT)}  \label{sec:btzerox0}
Consider the dynamical system  \eqref{eq:FOM} with zero initial conditions:
\begin{eqnarray}  \nonumber
\dot{\bx}(t) &=& \bA \bx(t)  + \bB \bu(t) ,\quad \bx(0) = 0, \\
 {\by}(t)  &=& \bC \bx(t).\label{eq:FOMu}
\end{eqnarray}
For simplicity of presentation, we assume that \eqref{eq:FOM} is \emph{controllable}  (characterized by the property that $\rank [s\bI-\bA, \bB]=n$ for all $s\in \mathbb C$)  and \emph{observable} ($\rank [s\bI-\bA^T, \bC^T]=n$ for all $s\in \mathbb C$).
The \emph{reachability Gramian} $\bP$ and the \emph{observability Gramian} $\bQ$ are the unique positive-definite solutions to the  Lyapunov equations
\begin{equation} \label{eq:lyap}
\bA \bP + \bP \bA^T + \bB \bB^T = 0\quad\mbox{and}\quad \bA^T \bQ + \bQ \bA + \bC^T \bC = 0,
\end{equation}
respectively.
Let $\bU$ and $\bL$ be Cholesky factors of $\bP$ and $\bQ$, respectively, i.e., $\bP = \bU \bU^T$ and
$\bQ = \bL \bL^T$. We note that in practice one computes $\bU$ and $\bL$ without ever forming $\bP$ and $\bQ$, see e.g., \cite{Ham82,SorZ03}. The method of \emph{balanced truncation} then computes the
singular value decomposition (SVD) of $\bU^T \bL= \bZ \Sigma \bY^T $ where $\Sigma= {\mathrm{diag}}( \sigma_1, \sigma_2 , \ldots , \sigma_n)$. The values $\{\sigma_i\}_1^n$ are the \emph{Hankel singular values} of \eqref{eq:FOM}. Only the part of the SVD associated with the large singular values above a given threshold value need be computed
with their associated left/right singular vectors.  This computation can also be performed without actually forming the product, $\bU^T \bL$.
Then, for a given reduction order $r$ and assuming for simplicity that $\sigma_r>\sigma_{r+1}$,  define
\begin{eqnarray} \label{bal_project}
\bV_{\mathsf{bt}} = \bU
 \bZ_r  \Sigma_r^{- 1/2} &{\rm and} &  \bW_{\mathsf{bt}}  = \bL
\bY_r \Sigma_r^{-1/2},
\end{eqnarray}
where $ \bZ_r $ and $\bY_r $ denote the leading $r$ columns of
$\bZ $, and $ \bY $, respectively, and
$\Sigma_r = {\mathrm{diag}}( \sigma_1, \sigma_2 , \ldots , \sigma_r)$.
Then, the order-$r$ reduced model  \eqref{eq:ROM} by balanced truncation is given by
\begin{equation} \label{eq:rombt}
 \Arbt = \Wbt^T \bA \Vbt, \quad \Brbt = \Wbt^T \bB, \quad \Crbt = \bC\Vbt .
 \end{equation}
The reduced model described by \eqref{eq:rombt}  is still asymptotically stable and
due to the  initial conditions, $\bx_0 = 0$, satisfies the error estimate
 \begin{equation} \label{eq:bterr}
 \left\| \by - \yrbt \right\|_{L_2} \leq \left(2\sum_{i=r+1}^n \sigma_i \right)  \left\| \bu\right\|_{L_2},
 \end{equation}
with $\left\| \bu \right\|_{L_2}:= (\int_0^\infty \bu(t)^T\bu(t)\, dt)^{1/2}$ denoting the norm in the function space $L_2^m$ of square Lebesgue integrable functions $u:[0,\infty)\to \mathbb R^m$.    Since the Hankel singular values can be computed consecutively, the decision about the choice of $r$ can be based on this error estimate \cite{GugA04}.
 \subsubsection{The Iterative Rational Krylov Algorithm (\IRKA)} \label{sec:irka}
In system \eqref{eq:FOMu} the output $\by(t)$ can be written as the convolution of {\it the impulse response}, defined as $\bh(t) = \bC e^{\bA t} \bB$  of \eqref{eq:FOMu}, with the forcing term $\bu(t)$; namely,
\begin{equation} \label{eq:u2y}
\by(t) =  \int_0^t \bh(t-\tau) \bu(\tau)  d \tau= \int_0^t \bC e^{\bA(t-\tau)} \bB \bu(\tau) d \tau,
\end{equation}
where $e^{\bA t}$ is the matrix exponential. The $L_2$ norm of the impulse response $\bh(t)$, see \cite{Ant05}, i.e.,
\begin{equation}
 \left\| \bh \right\|_{L_2}
 = \sqrt{\int_0^\infty \mathsf{trace}\left(\bh^T(t)\bh(t)\right) \, dt },
\end{equation}
is  the $\mathcal{H}_2$-norm of the underlying dynamical system.
The Iterative Rational Krylov Algorithm (\IRKA)~constructs a reduced model \eqref{eq:ROM} with the impulse response
$\tbh(t) = \tbC e^{\tbA t} \tbB$ such that the $\mathcal{H}_2$ norm of the error system is minimized; we seek $\tbh$ that solves the nonconvex $\mathcal{H}_2$ optimization problem ${\min_{\tbh(t)} \| \bh - \tbh \|_{L_2}}$.
 Since the computation of a global minimizer is difficult, one searches for a locally optimal reduced model that satisfies the first-order necessary conditions for optimality. These conditions can be formulated in terms of Lyapunov and Sylvester equations \cite{Wil90,HylB85} or alternatively, in terms of rational interpolation \cite{MeiL67,GugAB08,AntBG10}. It was shown in \cite{GugAB08} that the Lyapunov/Sylvester equation conditions and the rational interpolation conditions are theoretically equivalent. Interpolation-based optimality conditions make use of the \emph{transfer function} (characterized as the Laplace transform of the impulse response),  insofar as any (locally) $\mathcal{H}_2$-optimal reduced-order transfer function must be a (tangential) Hermite interpolant to the original transfer function.
\IRKA~is numerically efficient and only requires the solution of (sparse) shifted linear systems. For details on \IRKA~and more generally, optimal $\mathcal{H}_2$ approximation via rational interpolation, see \cite{GugAB08,AntBG10,BeaG15}.

For a given reduced order $r$, let
$\bV_{\mathsf{irka}} \in \IR^{n \times r}$ and $\bW_{\mathsf{irka}} \in \IR^{n \times r}$
denote the model reduction bases that \IRKA~produces.
Then, as in \BT, the reduced model \eqref{eq:ROM} by \IRKA, with the impulse response
${\tbh_\mathsf{irka}}(t) = \Crirka e^{\Arirka t} \Brirka$,
is obtained via projection
\begin{equation} \label{eq:romirka}
 \Arirka = \Wirkat \bA \Virka, \quad \Brirka = \Wirkat \bB, \quad \Crirka = \bC\Virka .
 \end{equation}
For the case of $\bx_0 = 0$,
the \IRKA~reduced model \eqref{eq:rombt}  satisfies
 \begin{equation} \label{eq:h2err}
 \left\| \by - \yrirka \right\|_{L_\infty} =  \sup_{t>0}\, \left\|\by(t)-\yrirka(t)\right\|_{\infty} \leq  \left\| \bh - {\tbh_\mathsf{irka}} \right\|_{L_2}
   \left\| \bu\right\|_{L_2}.
 \end{equation}

\subsection{Augmented \BT~for systems with nonhomogeneous initial conditions} \label{sec:reis}
As previously noted, the \BT~method does not take the initial condition, $\bx_0$, into account and so, the associated error bound is valid only for $\bx_0 = \mathbf{0}$.  As demonstrated in \cite{HeiRA11}, the resulting \BT-reduced model in \eqref{eq:rombt} may produce rather poor model reduction performance.  To overcome this difficulty, the authors of \cite{HeiRA11} proposed the following modification of \BT.

Suppose that the initial conditions of interest live in a subspace $\mathcal{X}_0$ spanned by the columns of a matrix $\bX_0 \in \IR^{n \times n_0}$. Constructing the augmented matrix $\bB_{\mathsf{aug}}  = [ \bB~\bX_0]$, then the \emph{augmented balanced truncation method (\ABT)} of  \cite{HeiRA11} applies the \BT~procedure of Section~\ref{sec:btzerox0} by replacing $\bB$ with
$\bB_{\mathsf{aug}}$; i.e., it applies \BT~to
\begin{eqnarray} \label{eq:FOMaug}
\dot{\bx}(t) = \bA \bx(t)  + [\bB~\bX_0] \bu_{\mathsf{aug}}(t), \quad
 {\by}(t)  = \bC \bx(t).
\end{eqnarray}
Denote by $\bP_{\mathsf{aug}}$ and $\bQ_{\mathsf{aug}}$, respectively, the reachability and observability Gramians of \eqref{eq:FOMaug}  and
let $\eta_i =  \sqrt{\lambda_i(\bP_{\mathsf{aug}}\bQ_{\mathsf{aug}})}$, for
$i=1,\ldots,n$, be the resulting Hankel singular values of \eqref{eq:FOMaug}.
Let $\bV_{{\mathsf{aug}}}$ and  $\bW_{{\mathsf{aug}}}$
 denote the corresponding \BT~model reduction projection spaces. Then, the reduced model of \cite{HeiRA11} is defined by
 \begin{equation} \label{eq:romaug}
 \Araug= \Waug^T \bA \Vaug, ~ \Braug = \Waug^T \bB, ~ \Craug = \bC\Vaug,
  ~\mbox{and}~
 \xraug = \Waug^T\bx_0.
 \end{equation}
Set $\Saug = {\mathsf{diag}}(\eta_1,\eta_2,\ldots,\eta_r)$ and let $\Qaug = \Laug \Laug^T$ be a Cholesky factorization of the observability Gramian of the augmented system \eqref{eq:FOMaug}.

If $\bx_0 = \bX_0 \bz_0$ and  $\widetilde{\bX}_0 = \Waug^T \bX_0$, then for the initial condition $\bx_0$ and the input $\bu(t)$, the output of the reduced model \eqref{eq:romaug} satisfies
the error estimate
 \begin{align}  \nonumber
 \left\| \by - \yraug \right\|_{L_2} &\leq \left(2\sum_{i=r+1}^n \eta_i \right)  \left\| \bu\right\|_{L_2} \\ &+ 3 \cdot 2^{-\frac{1}{3}}
 \left( \| \bL \bA \bX_0\|_2 + \| \Saug^{\frac{1}{2}} \Araug \widetilde{\bX}_0\|_2    \right)^{\frac{1}{3}} \left(\sum_{i=r+1}^n \eta_i \right)^{\frac{2}{3}} \| \bz_0\|_2,
 \label{eq:augerr}
 \end{align}
where $\| \cdot \|_2$ denotes the Euclidian vector norm, and the associated matrix norm, respectively.

Unlike standard \BT, the \ABT~approach offers an  error bound even in the case of nonzero initial conditions. This is a significant improvement compared to  regular \BT, but a potential issue with the error bound \eqref{eq:augerr} is that it depends on the reduced model to be computed. It is shown in \cite{HeiRA11} that if the augmented system \eqref{eq:FOMu} is {\it fully balanced}, which means that  the  Gramians are equal and diagonal, i.e. if $\bP = \bQ =\Sigma$, before reducing,
then the error bound can be simplified to
 \begin{align}  \nonumber
 \left\| \by - \yraug \right\|_{L_2} \leq & \left(2\sum_{i=r+1}^n \eta_i \right)  \left\| \bu\right\|_{L_2} \\ &+ 3
  \left\| \bSi^{\frac{1}{2}}_{{\mathsf{aug}}} \bA \right\|_2^{\frac{1}{3}} \Big\| {\bX}_0\Big\|_2^{\frac{1}{3}} \left(\sum_{i=r+1}^n \eta_i \right)^{\frac{2}{3}} \| \bz_0\|_2.
 \label{eq:augerrmod}
 \end{align}
However, since the transformation to fully balanced form may be numerically ill-conditioned, regular \BT~avoids this transformation by constructing the reduced balanced system directly, see \cite{Ant05} for details.

While the \ABT~method is clearly superior to regular \BT, it has some disadvantages as discussed in the introduction.  We will overcome these disadvantages with our new approach presented in the following section.

\section{A new model reduction method for systems with non-homogeneous initial conditions} \label{sec:main}

In this section we present a new flexible model reduction framework
for systems with nonzero initial conditions in which the map from input to output and the map from initial condition to output are reduced separately.

We discuss two approaches, the first approach discussed  in Section~\ref{sec:btbt} reduces both maps via \BT~and an upper bound for the approximation error in the output  is proved in Section~\ref{sec:boundbtbt}. In Section~\ref{sec:btirka}, motivated by the structure of the output error, we use the flexibility of the new framework and propose a combination of  \BT~and \IRKA.
\subsection{\BT~based reduction for $\Suy$ and $\Sxy$}  \label{sec:btbt}

Using the Duhamel formula for the system \eqref{eq:FOM}, the output  $\by(t)$ of \eqref{eq:FOM} can be explicitly written as
\begin{equation} \label{eq:yfom}
\by(t) = \underbrace{\bC e^{\bA t} \bx_0}_{\by_{x_0}(t)} + \underbrace{ \int_0^t \bC e^{\bA(t-\tau)} \bB \bu(\tau) d \tau}_{\byu},
\end{equation}
where $e^{\bA t}$ denotes the matrix exponential.   In \eqref{eq:yfom}, $\by_{x_0}(t)$ is the response of the system to the initial condition $\bx_0$ without any input, i.e., $\bu=0$, and $\byu$ is the response of the system to the input $\bu(t)$
with  zero initial conditions, i.e., $\bx_0=0$. Thus, due to the linearity of the underlying dynamics, the output is the superposition of these two signals. This fundamental observation indeed drives  our new approach, which in contrast to \ABT~does not mix the response of the initial condition with the response of the input response.

To derive the new approach in detail, recall that we have assumed that
\begin{equation}  \label{eq:z0}
\bx_0 = \bX_0 \bz_0
\end{equation}
for some $\bz_0 \in \IR^{n \times n_0}$, where the columns of $\bX_0$ form the basis for the subspace of relevant initial conditions. Then, $\by_{x_0}(t)$ can be re-written as
\begin{equation} \label{eq:yx0}
\by_{x_0}(t) = \bC e^{\bA t} \bX_0 \bz_0.
\end{equation}
This shows that $\by_{x_0}(t)$ is the response of a dynamical system
\begin{eqnarray} \nonumber
\dot{\bw}(t) &=& \bA \bw(t)  + \bX_0 \bv(t) ,\quad \bw(0) = 0, \\
 \by_{x_0}(t)  &=& \bC \bw(t),\label{eq:FOMx0}
\end{eqnarray}
with zero initial conditions and with input $\bv = \bz_0 \delta(t)$, where $\delta(t)$ denotes the Dirac delta distribution.  Thus, the approximation problem for $\by_{x_0}(t)$ (the response to an initial condition with zero forcing term) becomes a model reduction problem for a dynamical system with zero initial conditions.

Our new approach applies \BT~to
the dynamical system
\begin{eqnarray}  \nonumber
\dot{\bx}(t) &=& \bA \bx(t)  + \bB \bu(t) ,\quad \bx(0) = 0, \\
 {\byu}  &=& \bC \bx(t),\label{eq:FOMu2y}
\end{eqnarray}
as described in Section~\ref{sec:prep} and reduces its order to $r_\mathsf{u}$. Using the same notation as in Section~\ref{sec:prep}, the corresponding reduced model is denoted by
\begin{eqnarray}  \nonumber
\dot{\tbx}(t) &=& \Arbt \tbx(t)  + \Brbt \bu(t) ,\quad \bx(0) = 0, \\
\tbyu  &=& \Crbt \tbx(t). \label{eq:ROMu}
\end{eqnarray}
In parallel,  we apply \BT~to \eqref{eq:FOMx0} and reduce its order to $r_\mathsf{x_0}$.
Let  ${\bV}_{x_0} \in \IR^{n \times r_\mathsf{x0}}$ and
${\bW}_{x_0}\in \IR^{n \times r_\mathsf{x_0}}$ denote the corresponding \BT~projections. Then, the resulting reduced model
\begin{eqnarray} \nonumber
{\dot{\widetilde \bx}_{x_0}}(t) &=& \bA_{x_0} {\widetilde \bx}_{x_0}(t)  + {\widetilde \bX}_{0,x_0} \bv(t) ,\quad {\widetilde \bx}_{x_0}(0) = 0, \\
{\widetilde \by}_{x_0} &=& \bC_{x_0} {\widetilde \bx}_{x_0}(t),\label{eq:ROMx0}
\end{eqnarray}
has the coefficients
\begin{equation} \label{eq:romforx0}
 \bA_{x_0}= {\bW}_{x_0}^T \bA {\bV}_{x_0}, ~ {\widetilde \bX}_{0,x_0} = {\bW}_{x_0}^T \bX_0, ~\mbox{and}~ \bC_{x_0} = \bC{\bV}_{x_0}.
\end{equation}
Finally in order to approximate the complete output $\by(t)$ in \eqref{eq:yfom} with $\bx_0 = \bX_0 \bz_0$, we take the superposition of the output $\tbyu$ of \eqref{eq:ROMu} and the output ${\tilde \by}_{x_0}$ of \eqref{eq:ROMx0} with input $\bv(t) = \bz_0 \delta(t)$. This leads to the final approximation of $\by(t)$ by
 \begin{equation}  \label{eq:yrom}
\by(t) \approx {\widetilde \by}_{u,x_0}(t) = {\widetilde \by}_{x_0} + \tbyu= \underbrace{\bC_{x_0} e^{\bA_{x_0} t} {\widetilde \bX}_{0,x_0} \bz_0}_{{\widetilde \by}_{x_0}} + \underbrace{\int_0^t \Crbt e^{\Arbt(t-\tau)} \Brbt \bu(\tau) d \tau}_{\tbyu}.
 \end{equation}
Note that the two parts can be computed in parallel in an off-line phase
before using the reduced model for simulation, prediction, or in output control design.

A pseudocode of the method is presented in Algorithm~\ref{alg:us}.
\begin{algorithm}[hhh]
	\caption{\textcolor{black}{\BT-based model reduction for systems with nonzero initial conditions } \label{alg:us}}
	\vspace{1ex}
	\hspace{0.5ex}\textbf{Offline Phase:} \textsf{Construct the two reduced models}
	\begin{algorithmic}[1]
		\STATE \textbf{Input:} The system matrices $\bA$, $\bB$, $\bC$, and the initial condition basis $\bX_0$.
		\STATE \textbf{Output:} Reduced mappings for $\Suy$ and $\Sxy$.
	       \STATE  Approximating $\Suy:$ Apply $\BT$ to
	       $$
\Suy: \ \dot{\bx}(t) = \bA \bx(t)  + \bB \bu(t) ,\quad \bx(0) = 0,\qquad
 {\byu}  = \bC \bx(t)
	       $$
	       to obtain the reduced model $\tSuy$:
$$
\tSuy:  \ \dot{\tbx}(t) = \Arbt \tbx(t)  + \Brbt \bu(t) ,\quad \bx(0) = 0, \qquad
\tbyu  = \Crbt \tbx(t).  $$
	       \STATE  Approximating $\Sxy:$ Apply $\BT$ (or \IRKA,~see Section~\ref{sec:btirka}) to
	       $$
\Sxy: \
\dot{\bw}(t) = \bA \bw(t)  + \bX_0 \bv(t) ,\quad \bw(0) = 0,\qquad
 \by_{x_0}(t)  = \bC \bw(t)
	       $$
	       to obtain the reduced model
$$
\tSxy:\   {\dot{\widetilde \bx}_{x_0}}(t) = \bA_{x_0} {\widetilde \bx}_{x_0}(t)  + {\widetilde \bX}_{0,x_0} \bv(t) ,\quad {\widetilde \bx}_{x_0}(0) = 0, \qquad
{\widetilde \by}_{x_0} = \bC_{x_0} {\widetilde \bx}_{x_0}(t).  $$
	\end{algorithmic}	
		\vspace{1ex}
		\hspace{0.5ex}\textbf{Online Phase:}
		\textsf{Use  reduced models for simulation.~~}
	\begin{algorithmic}[1]	%
		\STATE \textbf{Input:} The initial condition $\bx_0$ and the forcing term $\bu(t)$.
		\STATE \textbf{Output:} The approximate reduced output ${\tilde \by}_{u,x_0}(t)$.
	       \STATE Compute $\bz_0$ such that 	$\bx_0 = \bX_0 \bz_0$.
		\STATE Simulate $\tSuy$ with input  $\bu(t)$ and {\it zero initial condition} to obtain	
		the output $\tbyu$.
	  \STATE Simulate $\tSxy$ with {\it zero input} and the initial condition $\bz_0$	
	to obtain	the output $\tbyx$.	
	\STATE Final approximate output: ${\widetilde \by}_{u,x_0}(t) = \tbyu + \tbyx$.
	\end{algorithmic}
\end{algorithm}

\subsection{Output error bounds for the proposed method} \label{sec:boundbtbt}
In this section, we establish an error bound for the approximation error $\| \by(t) - {\tilde \by}_{u,x_0}(t)\|$ in our new approach. For this we will employ the following result from \cite{Ant05}.
\begin{theorem}  \label{thm:aca}
Consider the dynamical system in \eqref{eq:FOMx0} and let
\begin{eqnarray} \nonumber
\dot{\bomega}(t) &=& \cbA \bomega(t)  + \cbB \bv(t) ,\quad \bomega(0) = 0, \\
 \by_{x_0}(t)  &=& \cbC \bomega(t),\label{eq:FOMx0bal}
\end{eqnarray}
be a fully balanced realization. For a given reduced order $r_\mathsf{x_0}$, partition
$\cbA$, $\cbB$ and $\cbC$ conformingly as
\begin{equation}
\cbA = \left[
\begin{array}{cc}
\cbA_{11} & \cbA_{12} \\ \cbA_{21} & \cbA_{22} \end{array}
\right],
\cbB = \left[
\begin{array}{cc}
\cbB_{1} \\ \cbB_{2}  \end{array}
\right],~\mbox{and}~
\cbC = \left[
\begin{array}{cc}
\cbC_{1} & \cbC_{2} \end{array}
\right],
\end{equation}
where $\cbA_{11} \in \IR^{r_\mathsf{x_0} \times r_\mathsf{x_0}}$,
$\cbB_{1} \in \IR^{r_\mathsf{x_0} \times m}$, and
$\cbC_{1} \in \IR^{p \times r_\mathsf{x_0}}$.
Moreover, let $\cbY \in \IR^{n \times r_\mathsf{x_0}}$ be a solution of the Sylvester equation
\begin{equation}
\cbA^T \cbY + \cbY \cbA_{11} + \cbC^T \cbC_1 = 0.
\end{equation}
Denote by $\bTheta = \mathsf{diag}(\theta_1,\theta_2,\ldots,\theta_n)$ the diagonal matrix of Hankel singular values
of \eqref{eq:FOMx0}, (or equivalently of \eqref{eq:FOMx0bal}) and
partition $\cbY$ and $\bTheta$ accordingly as
 \begin{equation}
 \cbY = \left[
\begin{array}{cc}
\cbY_{1} \\ \cbY_{2}  \end{array}
\right],\
\bTheta = \left[
\begin{array}{cc}
\bTheta_{1} &0 \\ 0 & \bTheta_{2} \end{array}
\right],
\end{equation}
where $\cbY_1, \bTheta_1 \in \IR^{r_\mathsf{x_0} \times r_\mathsf{x_0}}$. Let $\bhx(t)$ and $\tbhx(t)$ denote, respectively, the impulse response of the full-model
\eqref{eq:FOMx0} and the impulse response of its order-$r_{_\mathsf{x_0}}$ approximation \eqref{eq:ROMx0} by balanced truncation, then,
 \begin{eqnarray} \label{eq:aca}
\left\| \bhx -\tbhx\right\|_{L_2}^2  \leq \mathsf{trace}\left[ \left(\cbB_2 \cbB_2^T + 2 \cbY_2 \cbA_{12} \right) \bTheta_2 \right].
\end{eqnarray}
 \end{theorem}
As shown in \cite{Ant05}, the first term in the upper bound~\eqref{eq:aca}, i.e.,
$\mathsf{trace}\left[\cbB_2 \cbB_2^T  \bTheta_2\right]$ depends linearly on the neglected Hankel singular values
$\bTheta_2$, while the second term, $\mathsf{trace}\left[ 2 \cbY_2 \cbA_{12} \bTheta_2 \right]$ depends quadratically on
$\bTheta_2$.

Introducing, for simplicity of notation,
\begin{equation}  \label{eq:T}
\bT := \cbB_2 \cbB_2^T + 2 \cbY_2 \cbA_{12},
\end{equation}
we obtain the following upper bound for  the approximation error in our new approach.
\begin{theorem}  \label{thm:ub}
Let $\by(t)$ be the output of the full-model \eqref{eq:FOM} with initial condition $\bx_0$. Let ${\tilde \by}_{u,x_0}(t)$
be the reduced output obtained by Algorithm~\ref{alg:us}. Then
for any input $\bu(t)\in L_2^m$, the output error $\by(t) - {\tilde \by}_{u,x_0}(t)$ is bounded by
\begin{equation}  \label{eq:ub}
 \left\| \by - {\tilde \by}_{u,x_0} \right\|_{L_2} \leq \left(2\sum_{i=r_\mathsf{u}+1}^n \sigma_i \right)  \left\| \bu\right\|_{L_2} +
 \sqrt{ \mathsf{trace}\left[ \bT \,\bTheta_2 \right]}\, \left\| \bz_0 \right\|_2,
\end{equation}
where for $i=r_\mathsf{u}+1,\ldots,n$, $\sigma_i$  denotes the truncated Hankel singular values of \eqref{eq:FOMu2y}, for $i=r_{\mathsf{x_0}}+1,\ldots,n$, $\theta_i$  denotes
the truncated Hankel singular values of  \eqref{eq:FOMx0}, $\bT$ is as defined in \eqref{eq:T}, and
$\bz_0$ is as defined in  \eqref{eq:z0}.
\end{theorem}
\proof  Recall from \eqref{eq:yfom} that $\by = \by_{x_0}(t) + \byu$ and from  \eqref{eq:yrom} that   $\tby = {\tilde \by}_{x_0} + \tbyu$. Therefore,
\begin{equation} \label{eq:yminusyus}
 \left\| \by - {\tilde \by}_{u,x_0} \right\|_{L_2} \leq \left\| \byu - \tbyu \right\|_{L_2}  + \left\| \by_{x_0}(t) - {\tilde \by}_{x_0} \right\|_{L_2}.
\end{equation}
The first part of the upper bound in \eqref{eq:yminusyus} can be obtained by using the BT~upper bound  \eqref{eq:bterr}, since $\tbyu$ is the output of the reduced model approximation to
\eqref{eq:FOMu2y} obtained via \BT, and thus
\begin{equation}
\left\| \byu - \tbyu \right\|_{L_2} \leq  \left(2\sum_{i=r_\mathsf{u}+1}^n \sigma_i \right)  \left\| \bu\right\|_{L_2}.
\end{equation}
To bound the second part of the upper bound in \eqref{eq:yminusyus}, i.e., $\left\| \by_{x_0}(t) - {\tilde \by}_{x_0} \right\|_{L_2}$, we
use the definitions of $\by_{x_0}(t)$ and ${\tilde \by}_{x_0}$ in \eqref{eq:yx0} and  \eqref{eq:yrom}, respectively, to obtain
\begin{align} \label{eq:yxminusyxtilde}
 \left\| \by_{x_0}(t) - {\tilde \by}_{x_0} \right\|_{L_2} \leq  \left\| \bC e^{\bA t} \bX_0  - \bC_{x_0} e^{\bA_{x_0} t} {\widetilde \bX}_{0,x_0} \right\|_{L_2}  \left\| \bz_0 \right\|_2.
 \end{align}
Observe that $\bC e^{\bA t} \bX_0$ is the impulse response of \eqref{eq:FOMx0} and
$\bC_{x_0} e^{\bA_{x_0} t} {\widetilde \bX}_{0,x_0}$ is the impulse response of \eqref{eq:ROMx0}, which is obtained via \BT~of
\eqref{eq:FOMx0}.  Then, employing Theorem \ref{thm:aca} yields the desired result. $\Box$

The bound \eqref{eq:ub} on the output response error reveals the value of performing model reduction for the two mappings $\Suy$ and $\Sxy$ independently. First, observe that one can choose the reduced dimensions $r_\mathsf{u}$ and $r_\mathsf{x_0}$ independent of each other, 
based on given error tolerances or one can balance the model reduction errors resulting from reducing
$\Suy$ and $\Sxy$.  Secondly,  with independent reduction of $\Suy$ and $\Sxy$, scaling of input maps and initial conditions are uncoupled;  the resulting reduced models will be scale independent.

 In Section~\ref{sec:results} we will demonstrate by way of example that $\Sxy$ can be much easier to approximate than $\Suy$, suggesting that small values for $r_\mathsf{x_0}$ can suffice; or vice versa. The effect of performing the two reductions independently is reflected in the fact that the Hankel singular values $\{\sigma_i\}$ of $\Suy$  appear in the first term of \eqref{eq:ub} while the Hankel singular values $\{\theta_i\}$ of $\Sxy$ appear in the second term of
\eqref{eq:ub}.
This contrasts with the error bound \eqref{eq:augerrmod}, where the Hankel singular values of the augmented system will determine both components of the error.
More importantly, in \eqref{eq:augerrmod},  the diagonal matrix $\Saug$, consisting of the {\it retained} and {\it dominant} Hankel singular values, enters the upper bound. This is not desirable and is avoided by performing two model reduction steps independently. As \eqref{eq:ub} shows, in the new approach only the  truncated Hankel singular values $\{\sigma_{r_\mathsf{u}+1},\ldots,\sigma_n \}$  and $\{\theta_{r_\mathsf{x0}+1},\ldots,\theta_n \}$ occur in the  upper bound.

A significant advantage of the approach that we propose here is that different model reduction methods can be employed to compute independent reduced models for each of the maps $\Sxy$ and $\Suy$. This feature is exploited in the numerical examples of the following section to attain greater accuracy at smaller model order, but this can also be a significant advantage if one wishes to produce reduced models that retain special system structure inherited from the original system. For example, port-Hamiltonian systems may require that $\bB=\bC^T$ but this structure would be destroyed by  replacing $\bB$ with the augmented $[\bB~\bX_0]$.  In the approach we propose here, a structure-preserving model reduction method can be used for structured system $\Suy$, while an independent approach is used to reduce $\Sxy$.    A second example where this flexibility may become important is in the reduction of differential-algebraic systems, where initial conditions must be carefully selected to ensure consistency with the underlying constraint manifold.
This may also be reflected in the reduced model, requiring careful coordination of the reduced models produced for $\Sxy$ and $\Suy$.   Preservation of structure in reduced models and reduction of differential-algebraic systems considered with nonhomogeneous initial conditions are topics of active interest and will be pursued in later work.

\subsection{$\mathcal{H}_2$-based model reduction for $\Sxy$} \label{sec:btirka}
We now consider an alternative to the use of \BT~for reducing the map $\Sxy$. We can use \eqref{eq:yminusyus} and \eqref{eq:yxminusyxtilde} to express the bound for the error in approximating $\Sxy$ as
\begin{equation}
\left\| \by - {\tilde \by}_{u,x_0} \right\|_{L_2} \leq \underbrace{\left\| \byu - \tbyu \right\|_{L_2}}_{:=\,\eone}  +
 \underbrace{\left\| \bC e^{\bA t} \bX_0  - \bC_{x_0} e^{\bA_{x_0} t} {\widetilde \bX}_{0,x_0} \right\|_{L_2}}_{:=\,\etwo}  \left\| \bz_0 \right\|_2.
\end{equation}
From \eqref{eq:bterr}, \BT~is the appropriate model reduction technique to minimize $\eone$ in the above bound, just as we did in the previous section.  Employing \BT~also for the $\etwo$ component yields the bounds of Theorem~\ref{thm:aca}.  This combination leads to the error bound \eqref{eq:ub} where, based on a given tolerance and decay rate of Hankel singular values, we can determine the reduced dimensions $r_\mathsf{u}$ and $r_\mathsf{x_0}$ independently. In $\etwo$ the functions
$\bh_{x_0}(t) =\bC e^{\bA t} \bX_0 $ and ${\tilde \bh}_{x_0}(t) = \bC_{x_0} e^{\bA_{x_0} t} {\widetilde \bX}_{0,x_0}$ are the impulse responses of the full-model  \eqref{eq:FOMx0} and the reduced model \eqref{eq:ROMx0}, respectively. In other words,
$\etwo$ measures the $L_2$ error between the impulse responses $\bh_{x_0}(t)$ and ${\tilde \bh}_{x_0}(t)$; which is precisely  the $\mathcal{H}_2$ error norm as discussed in Section~\ref{sec:irka}. Therefore, it will be advantageous to approximate $\Sxy$ using model reduction techniques that aim to minimize the $\mathcal{H}_2$ error norm. Here we propose to use \IRKA~of \cite{GugAB08} in
 approximating $\Sxy$. The  corresponding change in Algorithm \ref{alg:us} will happen in Step 4 of the Offline Phase, where one would use \IRKA~instead of \BT.

The flexibility of being able to choose different methods for the individual problems, e.g., \BT~for $\eone$ and \IRKA~for $\etwo$,  highlights the advantage of solving the two model reduction problems separately.  Extensive numerical experience suggests that \IRKA~will generally produce smaller $\mathcal{H}_2$ errors than \BT \cite{GugAB08,AntBG10}.
This is an expected outcome to the extent that \IRKA~aims at producing a local minimizer to
 the $\mathcal{H}_2$ system error.  So, even though an error bound depending on Hankel singular values as in \eqref{eq:aca} is not known for \IRKA, we expect that approximating   $\Suy$ by \BT~and $\Sxy$ by \IRKA~leads to a smaller $L_2$ output error $\left\| \by - {\tilde \by}_{u,x_0} \right\|_{L_2}$.
This is demonstrated with some examples in the following section.

 \section{Numerical Results} \label{sec:results}
 In this section, we illustrate the theoretical discussion using two models, a  Mass-Spring-Damper Model and the ISS 1R Module.
For comparison, in each example we apply
\begin{itemize}[topsep=1ex,itemsep=-1ex,partopsep=1ex,parsep=1ex]
\item the augmented \BT~method of \cite{HeiRA11}, denoted by ``\textsf{AugBT}" in the plots and results,
\item Algorithm \ref{alg:us} with  \BT~in Step 4 of the Offline Phase, denoted by ``\textsf{BT-BT}", and
\item Algorithm \ref{alg:us} with \IRKA~in Step 4 of the Offline Phase
 \IRKA, denoted by ``\textsf{BT-IRKA}".
 \end{itemize}
 \subsection{Mass-Spring-Damper Model}
This is a slightly modified version of the coupled Mass-Spring-Spring model  in \cite[Sec 6.1]{GugPBS12}. We revised the model so that it
  has order $n=300$, i.e., a total of $150$ coupled
  mass-spring-dampers,   $m=10$ inputs and $p=1$ output. The inputs are the external forcing on the first $10$ masses and the output is the momentum of the first mass.  We choose a one-dimensional subspace for the initial condition $\bx_0$, i.e., $n_0=1$ and $\bX_0 \in \IR^n$.
In constructing $[\bB~\bX_0]$ for \textsf{AugBT}, $\bX_0$ is scaled appropriately as proposed in \cite{HeiRA11} so that the norm of the scaled initial condition  is equal to the maximum of the $2$-norms of the columns of $\bB$.

  \subsubsection{Case 1}
\begin{figure}[hhh]%
		\centering
\includegraphics[scale=.35]{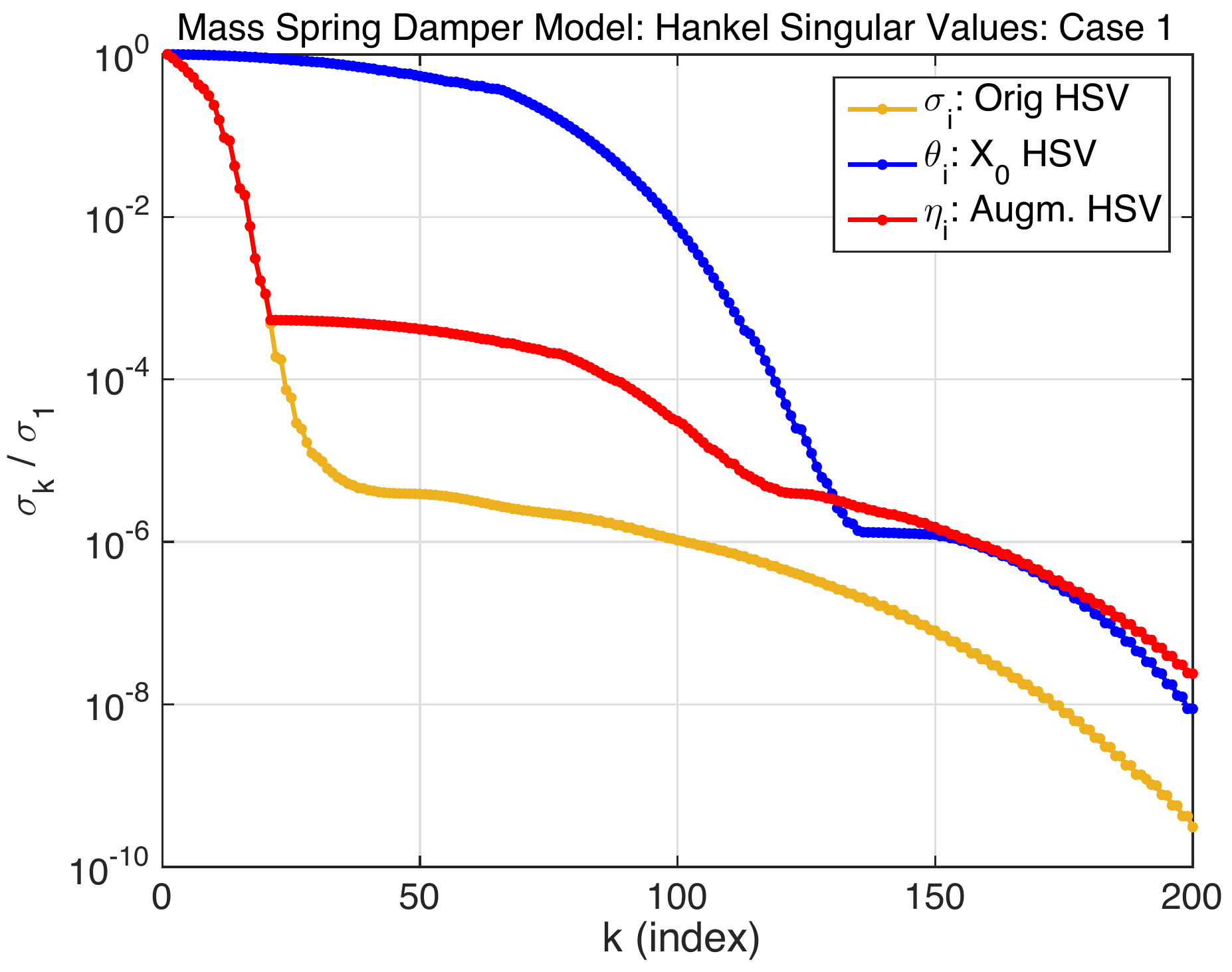}
\caption{Decay of the Hankel singular values (HSV)}
\label{fig:msdhsvcase1}
	\end{figure}

\begin{figure}
\centering
\begin{tabular}{cc}
\includegraphics[scale=.42]{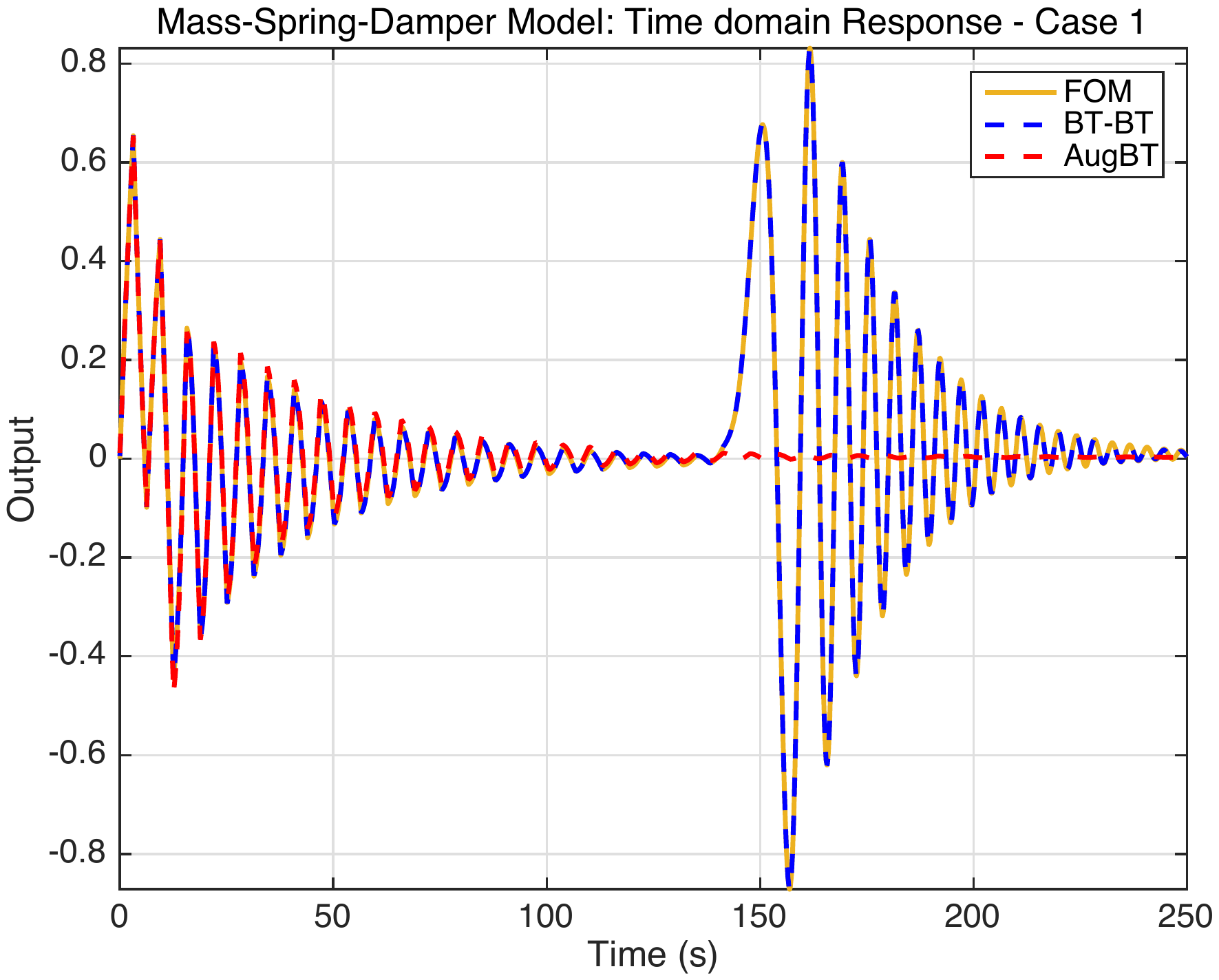}
&
\includegraphics[scale=.42]{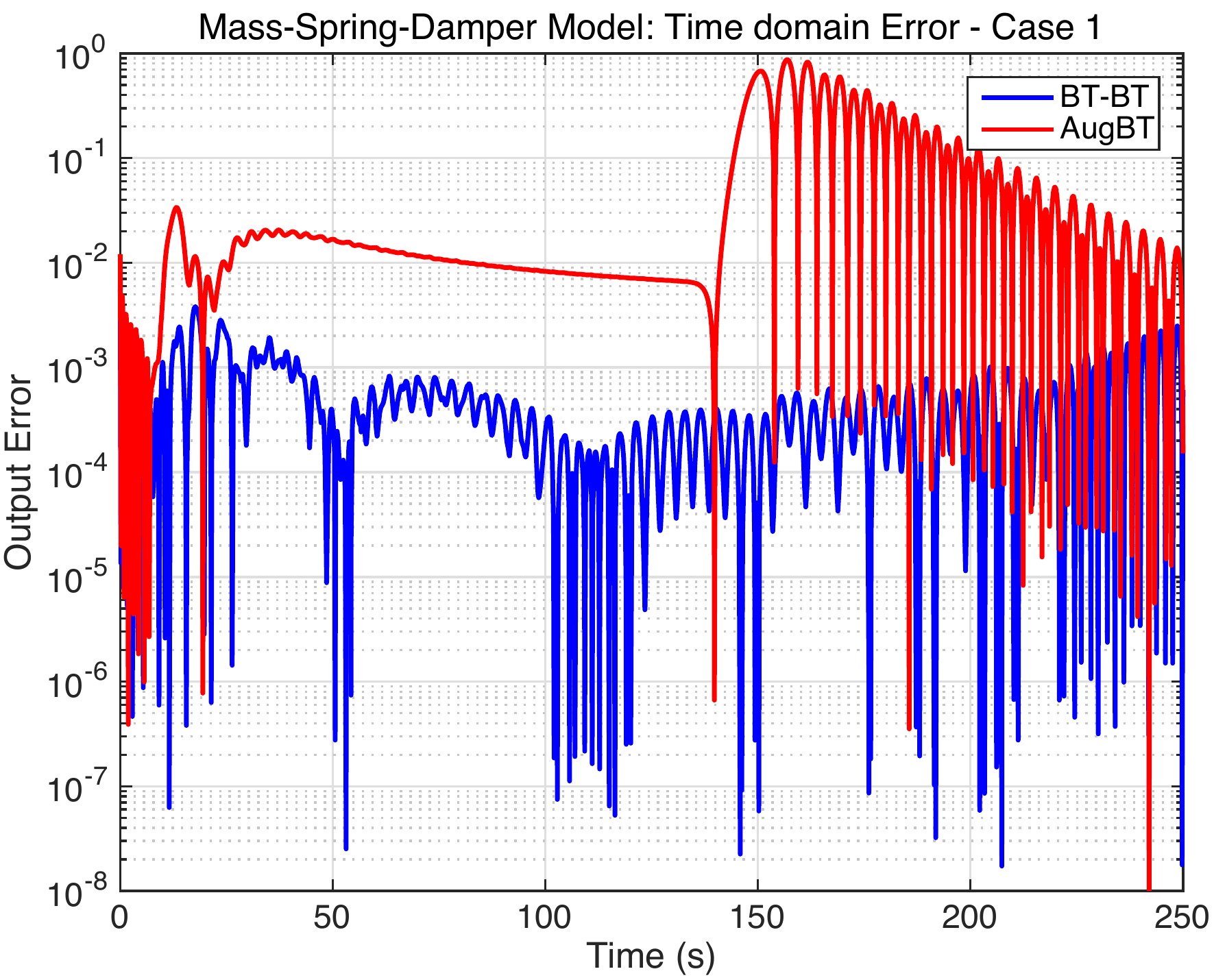}\\
\hspace*{0cm}{\scriptsize (a) Output Response} & \hspace*{0cm}{\scriptsize (b) Output Error }
\end{tabular}
\caption{Mass-Spring-Damper Model: Results for Case 1.}
\label{fig:msdresponsecase1}
\end{figure}

Choosing  $\bX_0 = \be_n$,  the $n^{\rm th}$ unit vector, the initial condition corresponds to the (variability in the) momentum of the $n^{\rm th}$ mass.  The Hankel singular values $\{\sigma_i\}$ of $\Suy$, and $\{\theta_i\}$ of
$\Sxy$, and $\{\eta_i\}$ of the augmented system $\mathcal{S}_\mathsf{aug}$ in  \eqref{eq:FOMaug}  are depicted in Figure \ref{fig:msdhsvcase1}. The figure illustrates that for this initial condition, the mapping $\Sxy$ has significantly slower decaying Hankel singular values than those of $\Suy$, i.e., $\Sxy$ is much harder to reduce than $\Sxy$. However, $\mathcal{S}_\mathsf{aug}$ misses this behavior and the decay of the leading $\eta_i$ follows almost exactly that of the $\sigma_i$; thus the difficulty in reducing $\Sxy$ is not visible.  We choose a truncation tolerance of $10^{-2}$ to automatically determine a reduced order based on the decay rate. In this case \textsf{AugBT} chooses $r_\textsf{aug}=16$. On the other hand, \textsf{BT-BT} chooses $r_\textsf{u} = 16$ and $r_{\mathsf{x_0}} = 98$. The important point here is that by performing model reduction separately, we are able to determine that  $\Sxy$ is very hard to reduce and an appropriate reduced order needs to be chosen for an accurate approximation.
To see the behavior of the reduced models, we simulated the full model and both reduced models for an exponentially decaying impulsive input for the nonzero initial condition case. The amplitudes of the input and input condition are chosen so that $\left\| \by_{\mathsf u} \right\|_{L_2} \approx \left\|\by_{x_0}\right\|_{L_2}$.
The output responses and the error in the output are presented in  Figure \ref{fig:msdresponsecase1}-(a) and Figure \ref{fig:msdresponsecase1}-(b), respectively. While \textsf{BT-BT} almost exactly replicates the full model response,
\textsf{AugBT} deviates from the response after approximately $t=140$  seconds. This delayed-response in the output  corresponds to the effect of the initial condition. Since the augmented singular values were lost in $\mathcal{S}_{\mathsf{aug}}$, \textsf{AugBT} cannot match this component. These results can be observed more clearly in the output error  in Figure \ref{fig:msdresponsecase1}-(b).
		
We also applied \textsf{BT-IRKA} and the results are very similar to (indeed better than) those obtained by \textsf{BT-BT}. However, to keep the figures easier to read, we have chosen not to include them in Figure \ref{fig:msdresponsecase1}. Instead in Table \ref{tab:msdcase1}, we list the relative $L_\infty$ and $L_2$ errors due to all three methods. As expected from the discussion in Section~\ref{sec:btirka}, performing model reduction on $\Sxy$ via  \IRKA~as opposed to \BT~reduces the output error even further.  As Figure \ref{fig:msdresponsecase1} already indicated, both \textsf{BT-BT} and \textsf{BT-IRKA} significantly outperform
\textsf{AugBT} for this example.
\begin{table}
\centering
\begin{tabular}{c||c|c|c}
& \textsf{AugBT} & \textsf{BT-BT}  & \textsf{BT-IRKA}   \\ \hline
$L_\infty$ error: &$ 9.9975 \times 10^{-1}$&$6.3534\times 10^{-3}$ & $5.7035 \times 10^{-3}$ \\
$L_2$ error:  & $8.8108 \times 10^{-1}$ & $3.7795 \times 10^{-3}$ &  $3.6008 \times 10^{-3}$
\end{tabular}	
\caption{Relative $L_\infty$ and $L_2$ errors in the output response for mass-spring-damper system, Case 1.}	
\label{tab:msdcase1}
\end{table}		
\subsubsection{Case 2}
\begin{figure}[hh]
\centering
\begin{tabular}{cc}
\includegraphics[scale=.40]{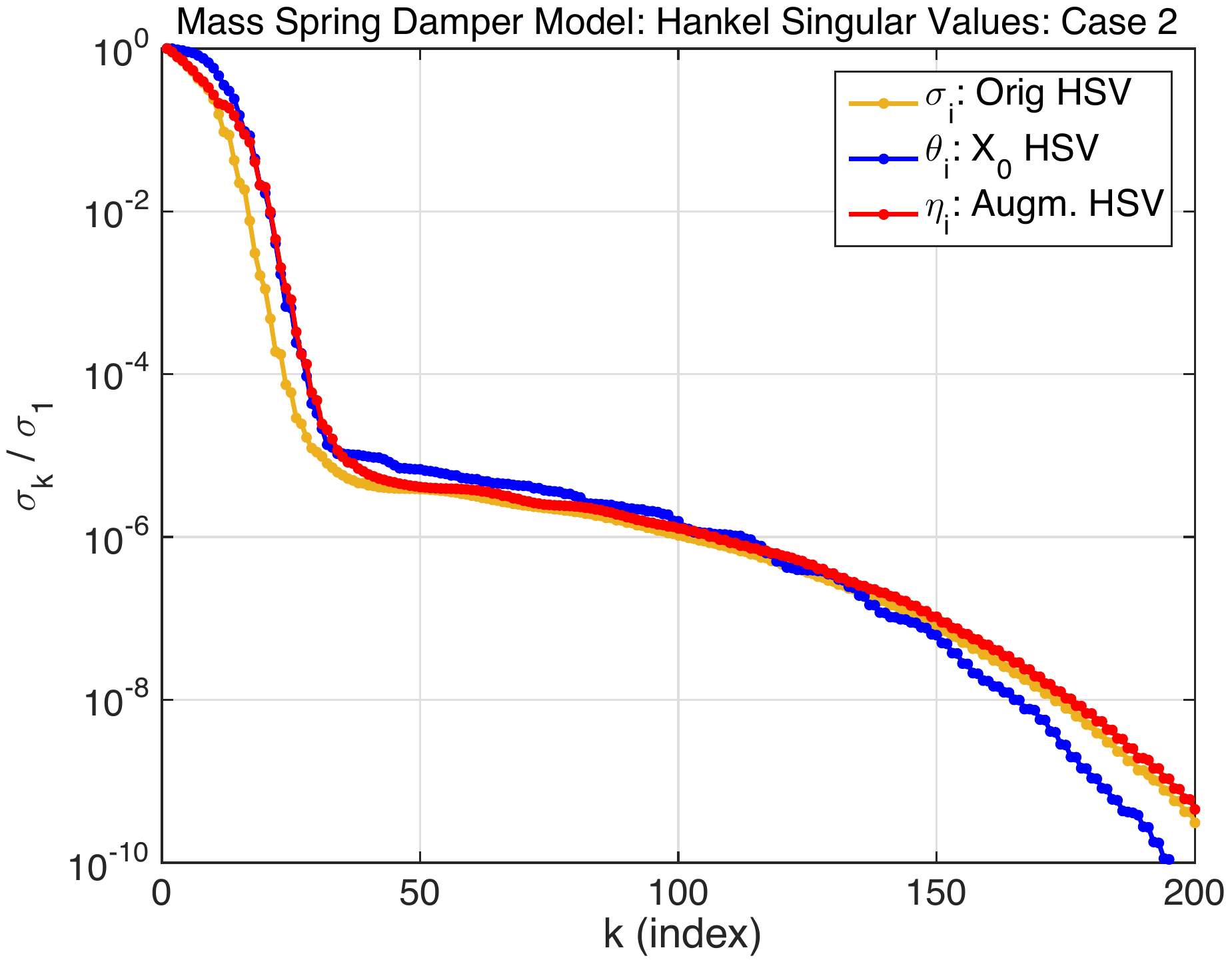}
&
\includegraphics[scale=.40]{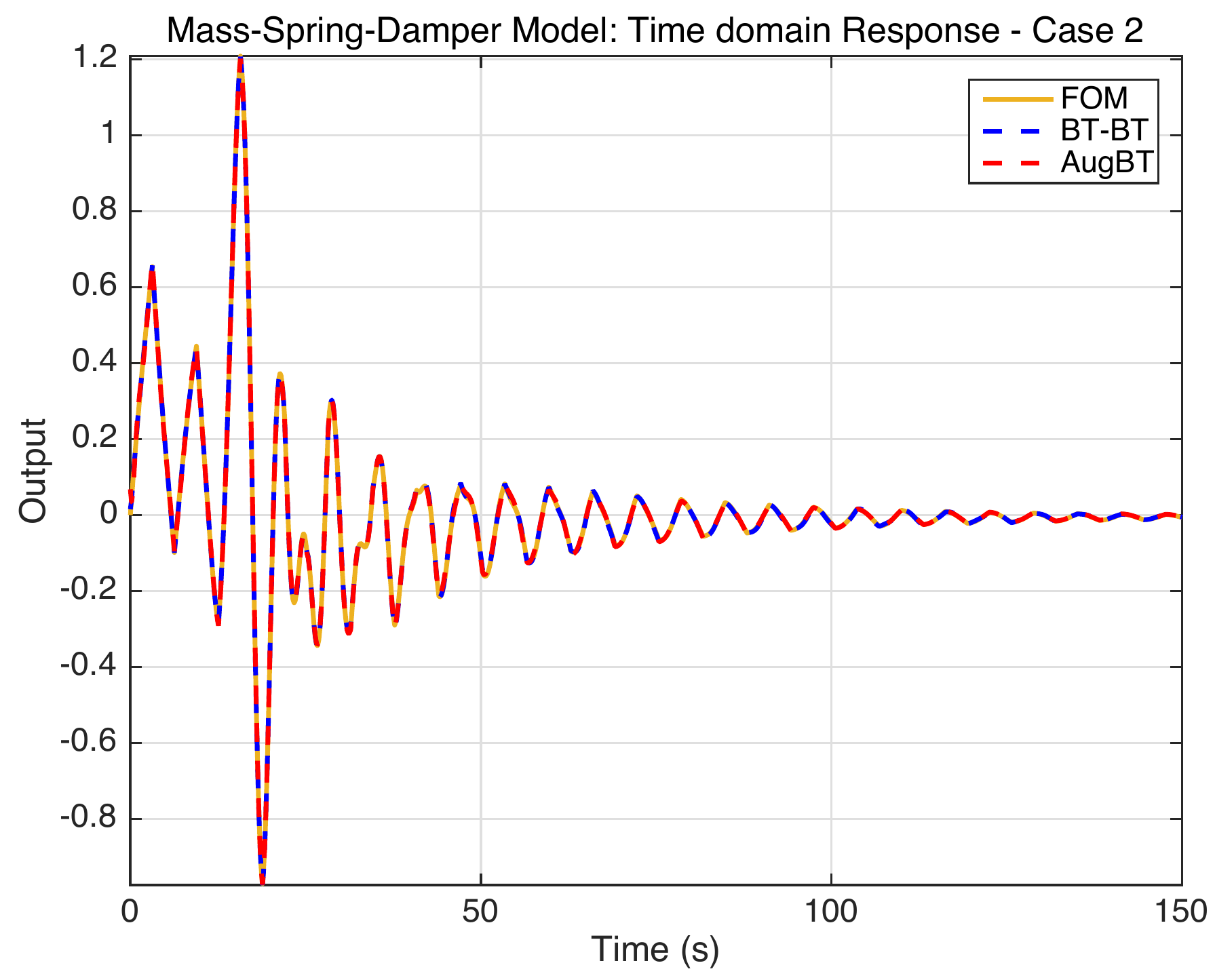}\\
\hspace*{0cm}{\scriptsize (a) Hankel Singular Values} & \hspace*{0cm}{\scriptsize (b) Output Response }
\end{tabular}
\caption{Mass-Spring-Damper Model: Case 2 Results}
\label{fig:msdcase2}
\end{figure}

For the second case, we change the initial condition to $\bX_0 = \be_{30}$ (corresponding to the momentum of the $15^{\rm th}$ mass) and repeat the experiments. The decay of the various Hankel singular values is shown in Figure \ref{fig:msdcase2}-(a); all showing a similar decay pattern unlike the previous case. The same tolerance of $10^{-2}$ as before yields
$r_\textsf{aug}=20$ for \textsf{AugBT}, and $r_\textsf{u} = 16$ and $r_{\mathsf{x_0}} = 20$
for \textsf{BT-BT}. The simulation results for the full-model and both reduced models given in Figure \ref{fig:msdcase2}-(b) illustrate that both \textsf{AugBT} and  \textsf{BT-BT} almost exactly replicates the true output.
As in Case 1, \textsf{BT-IRKA} performs even  better than \textsf{BT-BT} but is omitted in the figure. We present the relative $L_\infty$ and $L_2$ errors for all three methods in Table \ref{tab:msdcase2}. In terms of the $L_2$ norm, the performance of \textsf{AugBT} is close to that of \textsf{BT-BT}.  Performing \IRKA~on $\Sxy$ once again yields the best performance; not only in $L_2$ norm but also in the $L_\infty$ norm. Indeed, \textsf{BT-IRKA} is almost one order of magnitude better than
\textsf{AugBT} in the $L_\infty$ measure.
\begin{table}[hh]
\centering
\begin{tabular}{c||c|c|c}
& \textsf{AugBT} & \textsf{BT-BT}  & \textsf{BT-IRKA}   \\ \hline
$L_\infty$ error: &$5.6389\times 10^{-2}$ & $1.2753\times 10^{-2}$& $8.3127 \times 10^{-3}$ \\
$L_2$ error:  & $2.1969 \times 10^{-2}$ & $1.3017 \times 10^{-2}$ &  $6.4122 \times 10^{-3}$
\end{tabular}	
\caption{Relative $L_\infty$ and $L_2$ errors in the output response for mass-spring-damper system, Case 2}	
\label{tab:msdcase2}
\end{table}	
These two examples illustrate that even though \textsf{AugBT} can produce very good approximations, as in Case 2 above, the flexibility of performing model reduction separately can lead to big gains as in Case 1, where the decay of the augmented Hankel singular values has missed the initial condition formation. However, even in Case 2, where
	\textsf{AugBT} was successful, the proposed methods lead to  further improvements in accuracy, especially, when we employ \BT~on $\Suy$ and \IRKA~on	$\Sxy$.

\subsection{ISS IR Module}
In this section, we test the algorithms on the ISS 1R Module, one of the benchmark examples for model reduction, see \cite{GugAB01} for details. The model has order $n=270$, $m=3$ inputs and three outputs. For simplicity, we focus only on the first output, i.e., we take $p=1$.
 \subsubsection{Case 1}
\begin{figure}[hh]
\centering
\begin{tabular}{c}
\includegraphics[scale=.40]{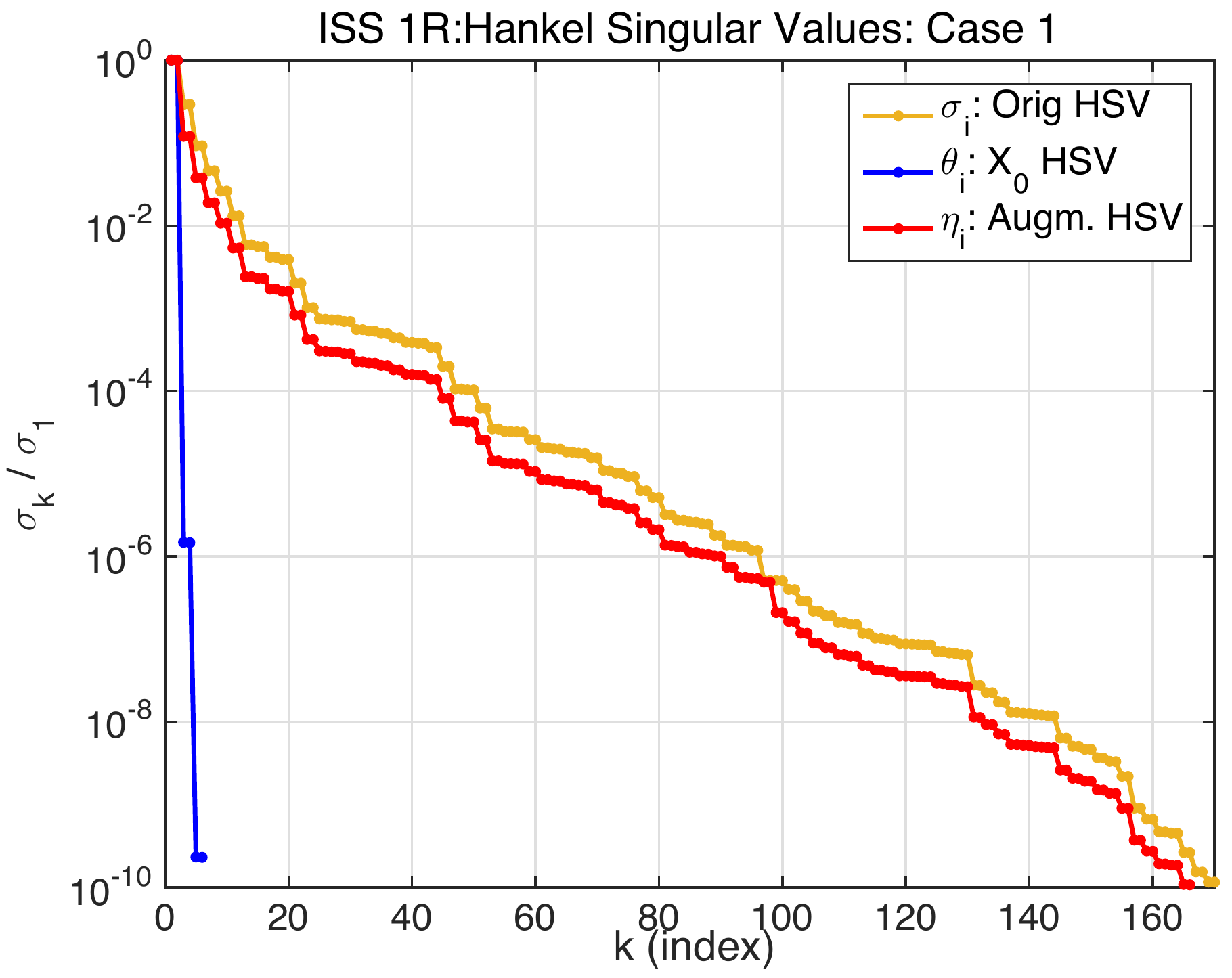}
\end{tabular}
\caption{ISS 1R Module: Hankel Singular Values}
\label{fig:isshsvdecay}
\end{figure}

As a first test case, we take the first three unit vectors as the $n_0=3$ dimensional basis $\bX_0$. The corresponding Hankel singular values are plotted in Figure \ref{fig:isshsvdecay}. The figure reflects that, unlike the previous example, $\Sxy$ is much easier to approximate than $\Suy$; indeed  $\Sxy$ has only $6$ nonzero Hankel singular values.
Using the truncation value of  $10^{-2}$, we obtain  $r_\mathsf{u} = 12$, $r_\mathsf{x0} = 2$, $r_\mathsf{aug} = 10$. The proposed approach, once again, makes the correct adjustment to the approximation order. We chose $\bx_0 = \alpha_1\be_2 + \alpha_2\be_3 \in \mathsf{span}(\bX_0)$,  exponential decaying sinusoidal forcing terms, and simulated the full and all three reduced models. Since all three reduced models yield very accurate approximations, the output response figures do not reveal much. Instead, we list the relative $L_\infty$ and $L_2$ errors in Table
\ref{tab:isscase1}. In this case \textsf{AugBT} and \textsf{BT-BT} perform very similarly in terms of the $L_2$ error;  \textsf{BT-BT}  is only slightly better. The difference in the $L_\infty$ performance is also comparable but once again favors  \textsf{BT-BT}. For this example,
\textsf{BT-IRKA} yields almost the same performance result as \textsf{BT-BT}. This is not surprising. The ratio $\frac{\theta_3}{\theta_1} = 1.4825 \times 10^{-6}$ reflects that the order $r_{\mathsf{x0}}=2$ approximation by \BT~and \IRKA~for on $\Sxy$ almost yields the same underlying minimal realization.
\begin{table}[hh]
\centering
\begin{tabular}{c||c|c|c}
& \textsf{AugBT} & \textsf{BT-BT}  & \textsf{BT-IRKA}   \\ \hline
$L_\infty$ error: &$ 6.6257\times 10^{-3}$ & $ 4.8068\times 10^{-3}$& $ 4.8068\times 10^{-3}$ \\
$L_2$ error:  & $ 1.8377 \times 10^{-3}$ & $1.6779 \times 10^{-3}$ & $1.6779 \times 10^{-3}$
\end{tabular}	
\caption{Relative $L_\infty$ and $L_2$ errors in output response for ISS 1R Module, Case 1}	 
\label{tab:isscase1}
\end{table}	
 \subsubsection{Case 2}
Using the same simulation as in the previous subsection we set $\bu(t) = 0$; thus only the initial condition drives the system. We compute the same measures as before and present them in Table~\ref{tab:isscase1nou}. All the reduced models are very accurate; however,  the new
method yields four orders of magnitude improvement  over \textsf{AugBT}. Note that since
$\bu(t)=0$,  \textsf{BT-BT} and  \textsf{BT-IRKA} are obtaining this accuracy using only a degree $r_{\mathsf{x0}}=2$ reduced model as opposed to \textsf{AugBT} where the reduced model has order $r_{\mathsf{aug}}=10$. This once again illustrates the flexibility of separating the two model reduction processes.
\begin{table}[hh]
\centering
\begin{tabular}{c||c|c|c}
& \textsf{AugBT} & \textsf{BT-BT}  & \textsf{BT-IRKA}   \\ \hline
$L_\infty$ error: &$ 8.1628\times 10^{-6}$ & $4.1782\times 10^{-11}$& $4.1782\times 10^{-11}$ \\
$L_2$ error:  & $7.9997 \times 10^{-6}$ & $3.1347 \times 10^{-10}$ &$3.1347 \times 10^{-10}$
\end{tabular}	
\caption{Relative $L_\infty$ and $L_2$ errors in output response for ISS 1R Module with $\bu(t)=0$}	
\label{tab:isscase1nou}
\end{table}

\section{Conclusions} \label{sec:conc}
We have described here a new approach for constructing a reduced order model for linear time invariant dynamical systems with nonhomogeneous initial conditions that originates with the observation that dynamical system response may be decomposed into an input-to-output map and an initial condition-to-output map which may then be treated independently of one another in creating an aggregate reduced order model for the full system. We have derived descriptive error bounds that improve upon other known bounds.  The advantages and flexibility of this new approach are demonstrated with a variety of numerical examples.

\bibliographystyle{plain}
\bibliography{InitalCondPaper}

\begin{thebibliography}{10}

\bibitem{Ant05}
A.C. Antoulas.
\newblock {\em Approximation of large-scale dynamical systems}.
\newblock Advances in Design and Control. Society for Industrial and Applied
  Mathematics, Philadelphia, PA, USA, 2005.

\bibitem{Ant05a}
A.C. Antoulas.
\newblock A new result on passivity preserving model reduction.
\newblock {\em Systems \& control letters}, 54(4):361--374, 2005.

\bibitem{AntBG10}
A.C. Antoulas, C.A. Beattie, and S.~Gugercin.
\newblock Interpolatory model reduction of large-scale dynamical systems.
\newblock In J.~Mohammadpour and K.~Grigoriadis, editors, {\em Efficient
  Modeling and Control of Large-Scale Systems}. Springer-Verlag, 2010.

\bibitem{BauBF14}
U.~Baur, P.~Benner, and L.~Feng.
\newblock Model order reduction for linear and nonlinear systems: a
  system-theoretic perspective.
\newblock {\em Archives of Computational Methods in Engineering},
  21(4):331--358, 2014.

\bibitem{BeaG11}
C.A. Beattie and S.~Gugercin.
\newblock Structure-preserving model reduction for nonlinear port-{H}amiltonian
  systems.
\newblock In {\em 50th IEEE Conference on Decision and Control and European
  Control Conference (CDC-ECC), 2011}, pages 6564--6569. IEEE, 2011.

\bibitem{BeaG15}
C.A. Beattie and S.~Gugercin.
\newblock Model reduction by rational interpolation.
\newblock In P.~Benner, A.~Cohen, M.~Ohlberger, and K.~Willcox, editors, {\em
  To appear in Model Reduction and Approximation: Theory and Algorithms.
  Available as http://arxiv.org/abs/1409.2140}. SIAM, Philadelphia, 2015.

\bibitem{CarFCA13}
K.~Carlberg, C.~Farhat, J.~Cortial, and D.~Amsallem.
\newblock The gnat method for nonlinear model reduction: effective
  implementation and application to computational fluid dynamics and turbulent
  flows.
\newblock {\em Journal of Computational Physics}, 242:623--647, 2013.

\bibitem{DrmGB14}
Z.~Drma\v{c}, S.~Gugercin, and C.A. Beattie.
\newblock Quadrature-based vector fitting for discretized $\mathcal{H}_2$
  approximation.
\newblock {\em SIAM J. Sci. Comp.}, 37(2):A625--A652, 2015.

\bibitem{Glo84}
K.~Glover.
\newblock All optimal {H}ankel-norm approximations of linear multivariable
  systems and their ${L}^{\infty}$-error bounds.
\newblock {\em Internat. J. Control}, 39(6):1115--1193, 1984.

\bibitem{GugA04}
S.~Gugercin and A.C. Antoulas.
\newblock A survey of model reduction by balanced truncation and some new
  results.
\newblock {\em Int. J. of Control}, 77:748--766, 2004.

\bibitem{GugAB08}
S.~Gugercin, A.C. Antoulas, and C.A. Beattie.
\newblock {$\mathcal{H}_2$ Model Reduction for Large-Scale Linear Dynamical
  Systems}.
\newblock {\em SIAM J. Matrix Anal. Appl.}, 30(2):609--638, 2008.

\bibitem{GugAB01}
S.~Gugercin, A.C. Antoulas, and N.~Bedrossian.
\newblock Approximation of the international space station 1r and 12a models.
\newblock In {\em IEEE Conference on Decision and Control}, volume~2, pages
  1515--1516, 2001.

\bibitem{GugPBS12}
S.~Gugercin, R.V. Polyuga, C.A. Beattie, and A.J.~{van der} Schaft.
\newblock Structure-preserving tangential interpolation for model reduction of
  port-{H}amiltonian systems.
\newblock {\em Automatica}, 48:1963--1974, 2012.

\bibitem{GusS99}
B.~Gustavsen and A.~Semlyen.
\newblock Rational approximation of frequency domain responses by vector
  fitting.
\newblock {\em IEEE Transactions on Power Delivery}, 14(3):1052--1061, 1999.

\bibitem{Ham82}
S.J. Hammarling.
\newblock Numerical solution of the stable, non-negative definite {L}yapunov
  equation.
\newblock {\em IMA Journal of Numerical Analysis}, 2(3):303--323, 1982.

\bibitem{HeiRA11}
M.~Heinkenschloss, T.~Reis, and A.C. Antoulas.
\newblock Balanced truncation model reduction for systems with inhomogeneous
  initial conditions.
\newblock {\em Automatica}, 47:559--564, 2011.

\bibitem{HylB85}
D.~Hyland and D.~Bernstein.
\newblock The optimal projection equations for model reduction and the
  relationships among the methods of {Wilson}, {Skelton}, and {Moore}.
\newblock {\em IEEE Transactions on Automatic Control}, 30(12):1201--1211,
  1985.

\bibitem{MayA07}
A.J. Mayo and A.C. Antoulas.
\newblock A framework for the solution of the generalized realization problem.
\newblock {\em Linear Algebra Appl.}, 425(2-3):634--662, 2007.

\bibitem{MeiL67}
L.~Meier~III and D.~Luenberger.
\newblock Approximation of linear constant systems.
\newblock {\em Automatic Control, IEEE Transactions on}, 12(5):585--588, 1967.

\bibitem{Moo81}
B.~Moore.
\newblock Principal component analysis in linear systems: Controllability,
  observability, and model reduction.
\newblock {\em IEEE Transactions on Automatic Control}, 26(1):17--32, 1981.

\bibitem{MulR76}
C.T. Mullis and R.~Roberts.
\newblock Synthesis of minimum roundoff noise fixed point digital filters.
\newblock {\em Circuits and Systems, IEEE Transactions on}, 23(9):551--562,
  1976.

\bibitem{PolS10}
R.V. Polyuga and A.J.~{van der} Schaft.
\newblock Structure preserving model reduction of port-{H}amiltonian systems by
  moment matching at infinity.
\newblock {\em Automatica}, 46:665--672, 2010.

\bibitem{Sor05}
D.C Sorensen.
\newblock Passivity preserving model reduction via interpolation of spectral
  zeros.
\newblock {\em Systems \& Control Letters}, 54(4):347--360, 2005.

\bibitem{SorZ03}
D.C Sorensen and Y.~Zhou.
\newblock Direct methods for matrix {S}ylvester and {L}yapunov equations.
\newblock {\em Journal of Applied Mathematics}, 2003(6):277--303, 2003.

\bibitem{Wil90}
D.A. Wilson.
\newblock Optimum solution of model-reduction problem.
\newblock {\em Proc. IEE}, 117(6):1161--1165, 1970.

\end{thebibliography}

\end{document}